# Quantitative Evaluation of Chaotic CBC Mode of Operation

Abdessalem Abidi[1], Qianxue Wang[3], Belgacem bouallègue[1], Mohsen Machhout[1] and Christophe Gyeux[2]

[1]Electronics and Microelectronics Laboratory
University of Monastir, Faculty of Sciences of Monastir,
Tunisia

[2] FEMTO-ST Institute, UMR 6174 CNRS DISC
Computer Science Department
University of Franche Comté, 16, Route de Gray, 25000, Besançon
France

[3]College of Automation
Guangdong University of Technology,
Guangzhou 510006
China
e-mail: abdessalemabidi9@gmail.com

*Abstract*—The cipher block chaining (CBC) block cipher mode of operation presents a very popular way of encrypting which is used in various applications. In previous research work, we have mathematically proven that, under some conditions, this mode of operation can admit a chaotic behavior according to Devaney. Proving that CBC mode is chaotic is only the beginning of the study of its security. The next step, which is the purpose of this paper, is to develop the quantitative study of the chaotic CBC mode of operation by evaluating the level of sensibility and expansivity for this mode.

## I. INTRODUCTION

Block ciphers have a very simple principle. They do not treat the original text bit by bit but they manipulate blocks of text for example, a block of 64 bits for the DES (Data Encryption Standard) or a block of 128 bits for the AES (Advanced Encryption Standard) algorithm. In fact, the original text is broken into blocks of N bits. For each block, the encryption algorithm is applied to obtain an encrypted block which has the same size. Then we gather all blocks, which are encrypted separately, to obtain the complete encrypted message. For decryption, we precede in the same way but this time starting from the cipher text to obtain the original message using the decryption algorithm instead of the encryption function. So, it is not sufficient to put anyhow a block cipher algorithm in a program. We can instead use these algorithms in various ways according to their specific needs. These ways are called the block cipher modes of operation. There are several modes of operation and each mode has owns characteristics and its specific security properties. In this article, we will consider only one of these modes, which is the cipher block chaining (CBC) mode.

The chaos theory we consider in this paper is the Devaney's topological one [1]. In addition to being recognized as one of the best mathematical definition of chaos, this theory offers a framework with qualitative and quantitative tools to evaluate the notion of unpredictability [2]. As an application of our fundamental results, we are interested in the area of information safety and security.

In this paper, which is an extension of our previous article [3], the theoretical study of the chaotic behavior for the CBC mode of operation is deepened by evaluating its level of sensibility and expansivity [4]. Our fundamental study is motivated by the desire to produce chaotic programs in the area of information security.

The remainder of this research work is organized as follows. In Section 2, we will recall some basic definitions concerning chaos and cipher-block chaining mode of operation. Section 3 is devoted to the results of our previous research works. In Section 4 quantitative topological properties for chaotic CBC mode of operation is studied in detail. This research work ends by a conclusion section in which our contribution is recalled and some intended future work are proposed.

## II. BASIC RECALLS

This section is devoted to basic definitions and terminologies in the field of topological chaos and in the one of block cipher mode of operation.

### A. Devaney's chaotic dynamical systems

In the remainder of this article, $S^n$ denotes the $n^{th}$ term of a sequence S while $\chi^{\mathbb{N}}$ is the set of all sequences whose elements belong to $\chi$. $V_i$ stands for the $i^{th}$ component of a vector V. $f^k = f \circ ... \circ f$ is for the $k^{th}$ composition of a function $f$. $\mathbb{N}$ is the set of natural (non-negative) numbers, while $\mathbb{N}^*$ stands for the positive integers 1, 2, 3, . . . Finally, the following notation is used: $[\![1; N]\!] = \{1, 2, ..., N\}$.

Consider a topological space $(\chi, \tau)$ and a continuous function $f: \chi \to \chi$ on $(\chi, \tau)$.

- *Definition 1*. The function $f$ is topologically transitive if, for any pair of open sets $U, V \subset \chi$ U, there exists an integer $k > 0$ such that $f^k(U) \cap V \neq \emptyset$.

- *Definition 2*. An element $x$ is a $peridic\ point$ for $f$ of period $n \in \mathbb{N}, n > 1, if\ f^n(x) = x$ and $f^k(x) \neq x$. $f$ is regular on $(\chi, \tau)$ if the set of periodic points for $f$ is dense in $\chi$ : for any point $x$ in $\chi$, any neighborhood of $x$ contains at least one periodic point.

- *Definition 3*. (Devaney's formulation of chaos [1]) The function $f$ is chaotic on $(\chi, \tau)$ if $f$ is regular and topologically transitive. The chaos property is strongly linked to the notion of "sensitivity", defined on a metric space $(\chi, \tau)$ by:

- *Definition 4*. The function $f$ has sensitive dependence on initial conditions if there exists $\delta > 0$ such that, for any $x \in \chi$ and any neighborhood $V$ of $x$, there exist $y \in V$ and $n > 0$ such that $d(f^n(x), f^n(y)) > \delta$. $\delta$ is called the constant of sensitivity of $f$.

Indeed, Banks et al. have proven in [5] that when $f$ is chaotic and $(\chi, \tau)$ is a metric space, then $f$ has the property of sensitive dependence on initial conditions (this property was formerly an element of the Devaney's definition of chaos). Additionally, the transitivity property is often obtained as a consequence of the strong transitivity one, which is defined below [6].

- *Definition 5*. $f$ is strongly transitive on $(\chi, d)$ if, for all point $x, y \in \chi$ and for all neighborhood $\sqrt{}$ of $x$, it exists $n \in \mathbb{N}$ and $x' \in \sqrt{}$ such that $f^n(x') = y$.

Finally, a function f has a constant of expansivity equal to $\varepsilon$ if an arbitrarily small error on any initial condition is always magnified until $\varepsilon$ [6]. Mathematically speaking,

Definition 6. The function f is said to have the property of *expansitivity* if $\exists \varepsilon > 0, \forall\ x \neq y, \exists\ n \in \mathbb{N}, d(f^n(x), f^n(y)) \geq \varepsilon$.

Then, $\varepsilon$ is the constant of expansivity of $f$. We also say that $f$ is $\varepsilon$-expansive.

B. CBC properties

Like some other modes of operation, the CBC mode requires not only a plaintext but also an initialization vector (IV) as input. In what follows, we will show how this mode of operation works in practice

*1) Initialisation vector IV*

As what have been already announced, in addition to the plaintext the CBC mode of operation requires an initialization vector in order to randomize the encryption. This vector is used to produce distinct cipher texts even if the same plaintext is encrypted multiple times, without the need of a slower re-keying process [7].

An initialization vector must be generated for each execution of the encryption operation, and the same vector is necessary for the corresponding execution of the decryption operation, see Figure 1. Therefore the IV, or information that is sufficient to calculate it, must be available to each party of any communication. The initialization vector does not need to be secret, so the IV, or information sufficient to determine the IV, may be transmitted with the cipher text. In addition, the initialization vector must be unpredictable: for any given plaintext, it must not be possible to predict the IV that will be associated to the plaintext, in advance to the vector generation [8].

There are two recommended methods for generating unpredictable IVs. The first method is to apply the forward cipher function, under the same key that is used for the encryption of the plaintext, to a nonce. The nonce must be a data block that is unique to each execution of the encryption operation.

For example, the nonce may be a counter or a message number. The second method is to generate a random data block using a FIPS (Federal Information Processing Standard)-approved random number generator [8, 9].

*2) Padding process*

A block cipher works on units of a fixed size (known as a block size), but messages come in variety of lengths. So some modes, namely the ECB (Electronic Codebook) and CBC ones, require that the final block is padded before encryption. In other words, the total number of bits in the plaintext must be a positive multiple of the block size N.

If the data string to be encrypted does not initially satisfy this property, then the formatting of the plaintext must entail an increase in the number of bits. A common way to achieve the necessary increase is to append some extra bits, called padding, to the trailing end of the data string as the last step in the formatting of the plaintext. An example of a padding method is to append a single 1 bit to the data string and then to pad the resulting string by as few 0 bits, possibly none, as are necessary to complete the final block (other methods may be used).

For the above padding method, the padding bits can be removed unambiguously provided the receiver can determine that the message is indeed padded. One way to ensure that the receiver does not mistakenly remove bits from an unpadded message is to require the sender to pad every message, including messages in which the final block is already complete. For such messages, an entire block of padding is appended. Alternatively, such messages can be sent without padding if, for each message, the existence of padding can be reliably inferred, e.g., from a message length indicator [8].

*3) CBC mode characteristics*

Cipher block chaining is a block cipher mode that provides confidentiality but not message integrity in cryptography. The CBC mode offers a solution to the greatest part of the problems presented by the ECB (Electronic codebook) for example [10], because thanks to CBC mode the encryption will depends on the context. Indeed, the cipher text of each encrypted block will depend not only on the initialization vector IV but also on the plaintext of all preceding blocks. Specifically, the binary operator XOR is applied between the current bloc of the plaintext and the previous block of the cipher text, as depicted in Figure 1. Then, we apply the encryption function to the result of this operation. For the first block, the initialization vector takes place of the previous cipher text block.

The main objective of this series of articles regarding the chaotic topological behavior of the CBC mode of operation is to understand in which extent this mode depends on its inputs. More precisely, is it possible to understand this dependence, in such a way that the effects of a modification of the IV and/or the message can be predicted? If so, this kind of weakness could be considered in the design of specific attacks, while if the converse is proven, that is to say, if the mid-to-long term effects of a slight modification of the input cannot be predicted, that chaotic dependence will make such attacks inefficient.

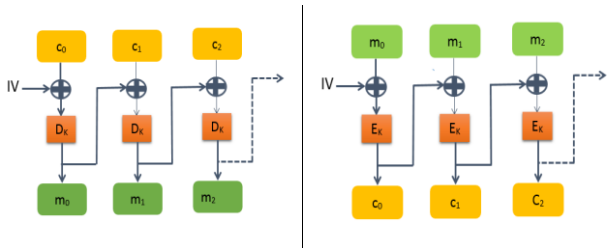

Fig. 1. CBC mode of operation

CBC mode has several advantages. In fact, this mode encrypts the same plaintext differently with different initialization vectors. In addition, the encryption of each block depends on the preceding block and therefore, if the order of the cipher text blocks is modified, the decryption will be impossible and the recipient realizes the problem. Furthermore, if a transmission error affects the encrypted block $C_i$, then only the blocks $m_i$ and $m_{i+1}$ are assigned, the other blocks will be determined correctly.

CBC has been the most commonly used mode of operation. Its main drawbacks are that encryption is sequential (i.e., it cannot be parallelized), and that the message must be padded to a multiple of the cipher block size. One way to handle this last issue is through the method known as cipher text stealing. Note that a one-bit change in a plaintext or IV affects all following cipher text blocks.

Decrypting with the incorrect initialization vector causes the first block of plaintext to be corrupted, but subsequent plaintext blocks will be correct. This is because a plaintext block can be recovered from two adjacent blocks of cipher text. As a consequence, decryption can be parallelized. Note that a one-bit change on the cipher text causes complete corruption of the corresponding block of plaintext, and inverts the corresponding bit in the following block of plaintext, but the rest of the blocks remain intact.

In what follows, we will summarize the results which proof the chaotic behavior of the CBC mode of operation.

### III. PREVIOUSLY OBTAINED RESULTS

In this section we will interest to our previous results which have been detailed in [3], in which we have been proven that some well chosen block ciphers can lead to a chaotically behavior for the CBC mode of operation. Indeed, this mode can be seen as a discrete dynamical system (recurrent sequence), whose evolution can thus be studied using common tools taken from the mathematical analysis [11].

#### A. Modeling the CBC mode as a dynamical system

Our modeling follows a same canvas than what has be done for hash functions [12, 6] or pseudorandom number generation [11].

Let us consider the CBC mode of operation with a keyed encryption function $E_k: B^N \rightarrow B^N$ depending on a secret key k, N is the size for the block cipher, and $D_k: B^N \rightarrow B^N$ is the associated decryption function, which is such that $\forall k, E_k \circ D_k$ is the identity function. We define the Cartesian product $X = B^N \times S_N$, where:

- B = {0, 1} is the set of Boolean values,
- $S_N = [\![2^N-1]\!]^N$ the set of infinite sequences of natural integers bounded by $2^N - 1$, or the set of infinite N-bits block messages, in such a way that $\chi$ is constituted by couples of internal states of the mode of operation together with sequences of block messages. Let us consider the initial function:

$$i: S_N \rightarrow [\![2^N-1]\!]^N$$
$$(m^i)_{i \in N} \rightarrow m^0$$

That the returns the first block of a (infinite) message, and the shift function:

$$S_N \rightarrow S_N$$
$$(m^0, m^1, m^2, \dots) \rightarrow (m^1, m^2, m^3, \dots)$$

which removes the first block of a message. Let $m_j$ be the $j$-th integer, or block message, $m \in [\![0, 2^N-1]\!]^N$, expressed in the binary numeral system, and when counting from the left. We define:

$$F_f: B^N \times [\![0, 2^N-1]\!]^N \rightarrow B^N$$
$$(x, m) \rightarrow (x_j m_j + f(x)_j \overline{m_j})_{j=1..N}$$

This function returns the inputted binary vector x, whose $m_j$-th components $x_{mj}$ have been replaced by $f(x)_{mj}$, for all $j = 1..N$ such that $m_j = 0$. In case where $f$ is the vectorial negation, this function will correspond to one XOR between the clair text and the previous encrypted state. So the CBC mode of operation can be rewritten as the following *dynamical system*:

$$\begin{cases} X^0 = (IV, m) \\ X^{n+1} = \left( E_k \circ F_{f_0} \left( i(X_1^N, X_2^N), \sigma(X_1^N) \right) \right) \end{cases}$$

For any given g: $[\![0, 2^N-1]\!]^N \times B^N \to B^N$, we denote $G_g(X) = (g(i(X_1), X_2); \sigma(X_1))$ (when $g = E_k \circ F_{f_0}$, we obtain one cipher block of the CBC, as depicted in Figure 1). So the recurrent relation of Eq.(1) can be rewritten in a condensed way, as follows:
$$X^{n+1} = G_{E_k \circ F_{f_0}}(X^n)$$
With such a rewriting, one iterate of the discrete dynamical system above corresponds exactly to one cipher block in the CBC mode of operation. Note that the second component of this system is a subshift of finite type, which is related to the symbolic dynamical systems known for their relation with chaos [13].

We now define a distance on χ as follows:
$d((x; m); (\check{x}; \tilde{m})) = d_e(x, \check{x}) + d_m(m, \tilde{m})$, where:

$$\begin{cases} d_e(x, \check{x}) = \sum_{k=1}^{N} \delta(x_k, \check{x}_k) \\ \\ d_m(m, \tilde{m}) = \frac{9}{N} \sum_{k=1}^{\infty} \frac{\sum_{i=1}^{N} |m_i - \tilde{m}_i|}{10^k} \end{cases}$$

This distance has been introduced to satisfy the following requirements:
- The integral part between two points $X$, $Y$ of the phase space χ corresponds to the number of binary components that are different between the two internal states $X_1$ and $Y_1$.
- The $k$-th digit in the decimal part of the distance between $X$ and $Y$ is equal to 0 if and only if the $k$-th blocks of messages $X_2$ and $Y_2$ are equal. This desire is at the origin of the normalization factor $\frac{9}{N}$.

### B. Proofs of chaos

As mentioned in Definition 3, a function $f$ is *chaotic* on (χ; τ) if $f$ is regular and topologically transitive.

We have began by stating some propositions that are primarily required in order to proof the chaotic behavior of the CBC mode of operation.

First of all, using the sequential characterization of the continuity, we have established that $G_g$ is a continuous map on $(\chi, d)$. After having recalled that a directed graph is strongly connected when it contains a directed path from u to v, for every pair of vertices u, v, we have established the following proposition.

**Proposition 1**: Let $g = E_k \circ F_{f_0}$, where is a given keyed block cipher and $f_0: B^N \to B^N$, $(x_1, ..., x_N) \to (\overline{x_1}, ..., \overline{x_N})$ is the Boolean vectorial negation. We consider the directed graph $G_g$, where:
- Vertices are all the N-bits words.
- There is an edge $m \in [\![0, 2^N - 1]\!]$ from $x$ to $\check{x}$ if and only if $g(m, x) = \check{x}$.

So if $G_g$ is strongly connected, then $G_g$ is strongly transitive.

Proving by doing so the strong transitivity of $G_g$ on $(\chi, d)$.

We have then proven that,

**Proposition 2:** If $G_g$ is strongly connected, then $G_g$ is regular.

According to Propositions 1 and 2, we can conclude that, if the directed graph $G_g$ is strongly connected, then the CBC mode of operation is chaotic according to Devaney, as established in our previous research work. In that article and for illustration purpose, we have also given some examples of encryption functions making this mode a chaotic one.

We have previously recalled that the mathematical framework of the theory of chaos offers tools to measure this unpredictable behavior quantitatively. The firsts of these measures are the constants of sensitivity and of expansivity, recalled in the definitions section. We now intend to investigate these measures.

### IV. QUANTITATIVE MEASURES

Let us firstly focus on the sensitivity property of the CBC mode of operation.

#### A. Sensitivity

**Property 1:** The CBC mode of operation is sensible to the initial condition, and its constant of sensibility is larger than the length N of the block size.

*Proof:* Let $X = (x; (m^0, m^1, ...)) \in X$ and $\delta > 0$. We are looking for $X' = (x'; (m'^0, m'^1, ...))$ and $n \in \mathbb{N}$ such that $d(X, X') < \delta$ and $d(G_g^n(X), G_g^n(X')) > N$.

Let us define $k_0 = \lfloor -\log_{10}(\delta) \rfloor + 1$, in such a way that all $X'$ of the form:
$$(X_1, (m^0, m^1, ..., m^{k_0}, m'^{k_0+1}, m'^{k_0+2}, ...))$$
are such that $d(X, X') < \delta$. In other words, all messages $m'$ whose $k_0$ first blocks are equal to $(m^0, m^1, ..., m^{k_0})$ are δ-close to $X$.

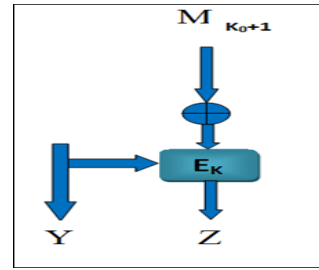

Fig. 2. $k_0+1$ th iterate of $G_g$

Let $y = G_g^{k_0}(X)_1$ and $z = G_g^{k_0+1}(X)_1$ as defined in Figure 2. We consider the block message $m'$ defined by:
$m' = y \oplus D_k(\bar{z})$ where $D_k$ is the keyed decryption function associated to $E_k$, and $\bar{z}$ is the negation of $z$. We thus define $X'$ as follow:
- $X'_1 = x$
- $\forall k \leq k_0, m'^k = m^k$
- $m'^{k_0+1} = m'$
- $\forall k \geq k_0 + 2, m'^k = \bar{m}^k$

So $d(G_g^{k_0+1}(X), G_g^{k_0+1}(X'))$
$= d(G_g(y; (m_{k_0+1}, m_{k_0+2}, ...)), G_g(y; (m', \overline{m_{k_0+1}}, \overline{m_{k_0+2}}, ...)))$

$$= d\left(\left(z;(m_{k_0+2}, m_{k_0+3}, \ldots)\right), \left(E_k(y \oplus m'); (\overline{m_{k_0+1}}, \overline{m_{k_0+2}}, \ldots)\right)\right)$$

$$= d\left(\left(z;(m_{k_0+2}, m_{k_0+3}, \ldots)\right), \left(E_k\left(y \oplus (y \oplus D_k(\bar{z}))\right); (\overline{m_{k_0+1}}, \overline{m_{k_0+2}}, \ldots)\right)\right)$$

$$= d\left(\left(z;(m_{k_0+2}, m_{k_0+3}, \ldots)\right), \left(E_k((y \oplus y) \oplus D_k(\bar{z})); (\overline{m_{k_0+1}}, \overline{m_{k_0+2}}, \ldots)\right)\right)$$

$$= d\left(\left(z;(m_{k_0+2}, m_{k_0+3}, \ldots)\right), \left(E_k(0 \oplus D_k(\bar{z})); (\overline{m_{k_0+1}}, \overline{m_{k_0+2}}, \ldots)\right)\right)$$

$$= d\left(\left(z;(m_{k_0+2}, m_{k_0+3}, \ldots)\right), \left(E_k(D_k(\bar{z})); (\overline{m_{k_0+1}}, \overline{m_{k_0+2}}, \ldots)\right)\right)$$

$$= d\left(\left(z;(m_{k_0+2}, m_{k_0+3}, \ldots)\right), \left(\bar{z}; (\overline{m_{k_0+1}}, \overline{m_{k_0+2}}, \ldots)\right)\right)$$

$$= d_e(z, \bar{z}) + d_m\left((m_{k_0+2}, m_{k_0+3}, \ldots), (\overline{m_{k_0+1}}, \overline{m_{k_0+2}}, \ldots)\right)$$

$$= N + \frac{9}{N} \sum_{k=k_0+2}^{\infty} \left(\frac{m_k - \overline{m_k}}{10^k}\right)$$

$$= N + \frac{9}{N} \sum_{k=k_0+2}^{\infty} \left(\frac{N}{10^k}\right)$$

$$= N + \frac{9}{N} \sum_{k=k_0+2}^{\infty} \left(\frac{1}{10^k}\right) = N + \frac{1}{10^{k_0+1}} > N,$$

which concludes the proof of the sensibility of $G_g$.

The second important tool which reinforces the chaotic behavior of the CBC mode of operation is the expansivity. The study of this property, which has been recalled in definition 6, will be regarded below.

### B. Expansivity

In this section we offer the proof that:

**Proposition 3:** *The CBC mode of operation is not expansive.*

*Proof:* consider for instance two initial vectors $x = (1,0,\ldots,0)$ and $x' = (0,1,0,\ldots,0)$, associated to the messages $m = ((0,1,0,\ldots,0), (0,\ldots,0), (0,\ldots,0), \ldots)$ and $m' = ((1,0,\ldots,0), (0,\ldots,0), (0,\ldots,0), \ldots)$: all blocks of messages are nul in both $m$ and $m'$, expect the first block. Let $X = (x, m)$ and $X' = (x', m')$.

Obviously, $x \neq x'$, while $x \oplus m_0 = x' \oplus m'_0$. This latter implies that $X_1^0 = X_1'^0$, and by a recursive process, we can conclude that $\forall i \in \mathbb{N}, X_1^i = X_1'^i$. So the distance between points $X = (x, m)$ and $X' = (x', m')$ is strictly positive, while for all $n > 0$ $d\left(G_g^n(X), G_g^n(X')\right) = 0$, which concludes the proof of the non expansive character of the CBC mode of operation by the mean of the exhibition of a counter example.

### V. CONCLUSION AND FUTURE WORK

In this paper, both expansivity and sensibility of symmetric ciphers are regarded, in the case of the CBC mode of operation. These quantitative topology metrics taken from the mathematical theory of chaos allow to measure in which extent a slight error in the initial condition is magnified during iterations. It is stated that, in addition to being chaotic as defined in the Devaney's formulation, the CBC mode of operation is indeed largely sensible to initial errors or modifications on either the IV or the message to encrypt. Its expansivity has been regarded too, but this property is not satisfied, as it has been established thanks to a counter example.

In future work, we intend to deepen the topological study of the behavior of the CBC mode of operation. We will study whether this mode of operation possesses other qualitative properties of disorder like the topological mixing. Additionally, other quantitative evaluations will be performed, and the level of topological entropy will be evaluated too.